\documentclass[preprint]{aastex}
\begin{document}
\title{Lyman-tomography of cosmic infrared background fluctuations with {\it Euclid}: probing emissions and baryonic acoustic oscillations at $z\gsim10$.}

\author{\ A. Kashlinsky\altaffilmark{1}, R.G. Arendt\altaffilmark{2}, F. Atrio-Barandela\altaffilmark{3}, and K. Helgason\altaffilmark{4}
\altaffiltext{1}{
Code 665, Observational Cosmology Lab, NASA Goddard Space Flight Center, 
Greenbelt, MD 20771 and
SSAI, Lanham, MD 20770; email: Alexander.Kashlinsky@nasa.gov} 
\altaffiltext{2}{
Code 665, Observational Cosmology Lab, NASA Goddard Space Flight Center, 
Greenbelt, MD 20771 and
UMBC, Baltimore, MD 21250} 
\altaffiltext{3}{Dept of Theoretical Physics, University of Salamanca, Spain.
} 
\altaffiltext{4}{MPA, Karl-Schwarzschild-Str. 1, 85748 Garching, Germany}
}


\def\plotone#1{\centering \leavevmode
\epsfxsize=\columnwidth \epsfbox{#1}}

\def\wisk#1{\ifmmode{#1}\else{$#1$}\fi}

\def\wm2sr {Wm$^{-2}$sr$^{-1}$ }		
\def\nw2m4sr2 {nW$^2$m$^{-4}$sr$^{-2}$\ }		
\def\nwm2sr {nWm$^{-2}$sr$^{-1}$\ }		
\def\nw2m4sr {nW$^2$m$^{-4}$sr$^{-1}$\ }
\def\Ncut {$N_{\rm cut}$\ }
\def\lt     {\wisk{<}}
\def\gt     {\wisk{>}}
\def\le     {\wisk{_<\atop^=}}
\def\ge     {\wisk{_>\atop^=}}
\def\lsim   {\wisk{_<\atop^{\sim}}}
\def\gsim   {\wisk{_>\atop^{\sim}}}
\def\kms    {\wisk{{\rm ~km~s^{-1}}}}
\def\Lsun   {\wisk{{\rm L_\odot}}}
\def\Msun   {\wisk{{\rm M_\odot}}}
\def\um     { $\mu$m\ }
\def\sig    {\wisk{\sigma}}
\def\etal   {{\sl et~al.\ }}
\def\eg	    {{\it e.g.\ }}
\def\ie     {{\it i.e.\ }}
\def\bsl    {\wisk{\backslash}}
\def\by     {\wisk{\times}}
\def\cosec {\wisk{\rm cosec}}
\def\mic {\wisk{ \mu{\rm m }}}

\def\amin   {\wisk{^\prime\ }}
\def\asec   {\wisk{^{\prime\prime}\ }}
\def\cc     {\wisk{{\rm cm^{-3}\ }}}
\def\deg     {\wisk{^\circ}}
\def\ddeg   {\wisk{{\rlap.}^\circ}}
\def\damin  {\wisk{{\rlap.}^\prime}}
\def\dasec  {\wisk{{\rlap.}^{\prime\prime}}}
\def\approxeq{$\sim \over =$}
\def\abouteq{$\sim \over -$}
\def\percm{cm$^{-1}$}
\def\percmsq{cm$^{-2}$}
\def\percmcub{cm$^{-3}$}
\def\perhz{Hz$^{-1}$}
\def\perpetratorsc{$\rm pc^{-1}$}
\def\persec{s$^{-1}$}
\def\peryr{yr$^{-1}$}
\def\te{$\rm T_e$}
\def\tenup#1{10$^{#1}$}
\def\to{\wisk{\rightarrow}}
\def\thin{\thinspace}
\def\uk{$\rm \mu K$}
\def\p{\vskip 13pt}
\begin{abstract}
The {\it Euclid} space mission, designed to probe evolution of the Dark Energy, will map a large area of the sky at three adjacent near-IR filters, Y, J and H. This coverage will also enable mapping source-subtracted cosmic infrared background (CIB) fluctuations with unprecedented accuracy on sub-degree angular scales. Here we propose methodology, using the Lyman-break tomography applied to the {\it Euclid}-based CIB maps, to accurately isolate the history of CIB emissions as a function of redshift from $10\lsim z \lsim 20$, and to identify the baryonic acoustic oscillations (BAOs) at those epochs. To identify the BAO signature, we would assemble individual CIB maps over conservatively large contiguous areas of $\gsim 400$deg$^2$. The method can isolate the CIB spatial spectrum by $z$ to  sub-percent statistical accuracy. We illustrate this with a specific model of CIB production at high $z$ normalized to reproduce the measured {\it Spitzer}-based CIB fluctuation. We show that even if the latter contain only a small component from high-$z$ sources, the amplitude of that component can be accurately isolated with the methodology proposed here and the BAO signatures at $z\gsim 10$ are recovered well from the CIB fluctuation spatial spectrum. Probing the BAO at those redshifts will be an important test of the underlying cosmological paradigm, and would narrow the overall uncertainties on the evolution of cosmological parameters, including the Dark Energy. Similar methodology is applicable to the planned {\it WFIRST} mission, where we show that a possible  fourth near-IR channel at $\geq 2\mu$m would be beneficial.
\end{abstract}
\section{Introduction}
Cosmic Infrared Background (CIB) contains emissions from first sources at the end of the ``Dark Ages", individually inaccessible to telescopic studies (see review by Kashlinsky 2005). Significant development in identifying CIB fluctuations from early times came with the discovery of source-subtracted CIB fluctuations in deep {\it Spitzer} data (Kashlinsky et al. 2005, 2007a, 2012 - KAMM1, KAMM2, K12)
which strongly exceed fluctuations from remaining known galaxies (KAMM1, Helgason et al. 2012 - HRK12).  It was suggested that these fluctuations arise at epochs associated with the first-stars era (KAMM1, Kashlinsky et al. 2007b - KAMM3)
or in yet undiscovered populations at low $z$, ripped off from their galaxies and contributing the intrahalo light (Cooray et al. 2012). 

{\it Euclid} (Laureijs et al. 2011) is designed to probe evolution of the ``dark energy" (DE) 
and provide a near-IR coverage over a substantial part of the sky. The instrumentational and observational characteristics of the mission make it uniquely suitable for the near-IR CIB measurements, which this team will perform via a NASA-funded project LIBRAE (Looking at Infrared Background Radiation Anisotropies with Euclid). This {\it Letter} shows how
tomographic analysis of the {\it Euclid} data, 
using the Lyman-break feature in the portion of the CIB from pre-reionization epochs, can 1) isolate CIB contributions as a function of $z$ at $10\lsim z \lsim 20$, and 2) probe the BAOs at those epochs.  

\section{Methodology and application to {\it Euclid} parameters}

The spectral energy distribution (SED) of sources at high $z$ exhibits a cutoff at energies above the Lyman limit (e.g. Haardt \& Madau 2012). In the presence of significant amounts of neutral hydrogen (\ion{H}{1}) such cutoff would lie at the Ly$\alpha$ transition of 10.2 eV (0.122 \mic), while if the surrounding hydrogen were ionized (\ion{H}{2}) the cutoff would be likely at the Ly-continuum of 13.6 eV (0.0912 \mic). Because of the cutoff in the SED of populations, a filter at $\lambda$ sees only sources at $z\leq z_{\rm Ly}(\lambda)\equiv \frac{\lambda}{\lambda_{\rm Ly-break}}-1$ with $\lambda$ always corresponding to the {\it longest} wavelength of the filter. $z_{\rm Ly}$  may vary by $\sim20\%$ due to the Lyman-continuum vs Ly$\alpha$ cutoffs.  Observations of the Gunn-Peterson absorption suggest presence of \ion{H}{1} at $z\gsim 6-7$, making it likely that at $z\gsim 10$ the cutoff in the SED of the  objects lies at Ly$\alpha$ (Djorgovski et al. 2003, and references therein).

After Fourier transforming CIB fluctuations, $f($\mbox{\boldmath$q$}$)= \int \delta
F($\mbox{\boldmath$x$}$)
\exp(-i$\mbox{\boldmath$x$}$\cdot$\mbox{\boldmath$q$}$) d^2x$, the (auto-)power spectrum at $\lambda_1$ is $P_1(q)=\langle |
f($\mbox{\boldmath$q$}$)|^2\rangle$, with the average taken over
the independent Fourier elements corresponding to the given $q$. The cross-power  between fluctuations at  $\lambda_1,\lambda_2$ is
$P_{\rm 12} (q)$=$\langle f_{1}(q) f^*_{2}(q)\rangle$.  
The coherence between the two bands is ${\cal C}_{12}\equiv \frac{P_{12}^2}{P_1P_2}\leq1$. The mean square fluctuation 
on angular scale $2\pi/q$ is $\frac{q^2P}{2\pi}$ and the cyclical wavenumber $q$ is related to multipole $\ell\simeq q$ (in radian$^{-1}$). 

 The projected CIB auto-power 
is related to the underlying 3-D power, $P_{3D}$, of the sources by the relativistic Limber equation: $P_{\lambda}(q)= \int
 (\frac{dF_{\lambda^\prime}}{dz})^2 Q(qd_A^{-1}; z) dz$, where $d_A$ is the comoving angular distance to $z$,  $Q(k, z) \equiv \frac{P_{3D}(k, z)}{c(1+z)dt/dz d_A^2(z)}$ and $\frac{dF_{\lambda^\prime}}{dz}$ is the CIB flux production at rest $\lambda^\prime\equiv\lambda/(1+z)$ over the epochs spanned by the integration. Assuming a flat Universe with matter, DE,  radiation/relativistic component and curvature density parameters: $\Omega_m, \Omega_{\rm DE}, \Omega_\gamma, \Omega_k$, leads to
 $c(1+z)dt/dz = cH_0^{-1}/[E(z)]^\frac{1}{2}$ and $d_A(z)=\int_0^z dz^\prime/E(z^\prime)$, where $E(z)\equiv \Omega_\gamma(1+z)^4+ \Omega_m(1+z)^3+\Omega_k(1+z)^2+\Omega_{\rm DE}f(z)$, with $f(z)$ describing the $z$-evolution of DE. Then the mean squared flux fluctuation at  $\lambda$ can be rewritten as:
\begin{equation}
\frac{q^2P_{\lambda}(q,<z_{\rm Ly}(\lambda))}{2\pi}= \int_0^{z_{\rm Ly}(\lambda)}
 \left(\frac{dF_{\lambda^\prime}}{dz}\right)^2 \Delta^2(qd_A^{-1}; z) E(z)dz
\label{eq:limber_1}
\end{equation}
where $\Delta^2(k,z)\equiv \frac{k^2P_{3D}(k,z)}{2\pi cH_0^{-1}}$ is the mean squared fluctuation in the source counts over a cylinder of diameter $k^{-1}$ and 
length $R_H\equiv cH_0^{-1}$ (Kashlinsky 2005).  {\it The integration range 
stops at} $z_{\rm Ly}(\lambda)$ because sources at larger redshifts emit only longward of $\lambda_{\rm Ly-break}$,
corresponding to 
the {\it far} edge of the filter of band $\lambda$. 
The cross-power between two bands, $\lambda_2>\lambda_1$, extends only to
$z_{\rm Ly}(\lambda_1)$:
\begin{equation}
P_{12} = \int_0^{z_{\rm Ly}(\lambda_1)}
 \frac{dF_{\lambda_1^\prime}}{dz} \frac{dF_{\lambda_2^\prime}}{dz} Q(qd_A^{-1}; z) dz.
\label{eq:cross-power}
\end{equation}
For $\lambda_2>\lambda_1$, we write:
\begin{equation}
P_{2}  (q,<z_{\rm Ly}(\lambda_2))= \int_{z_{\rm Ly}(\lambda_1)}^{z_{\rm Ly}(\lambda_2)} 
\left(\frac{dF_{\lambda_2^\prime}}{dz}\right)^2 Q(qd_A^{-1}; z) dz
\;\;+P_{2}(q, <z_{\rm Ly}(\lambda_1)) = P_{\Delta z} + \frac{1}{{\cal C}_{12}(z<z_{\rm Ly}(\lambda_1))}\frac{P_{12}^2}{P_1}.
\label{eq:limber_2}
\end{equation}
$P_{\Delta z}$ above probes emissions spanning $\Delta z$ at $z_{\rm Ly}(\lambda_1)<z<z_{\rm Ly}(\lambda_2)$ and arises from populations inaccessible to $\lambda_1$, but present at $\lambda_2$.  

\begin{figure}[h!]
\includegraphics[width=7in]{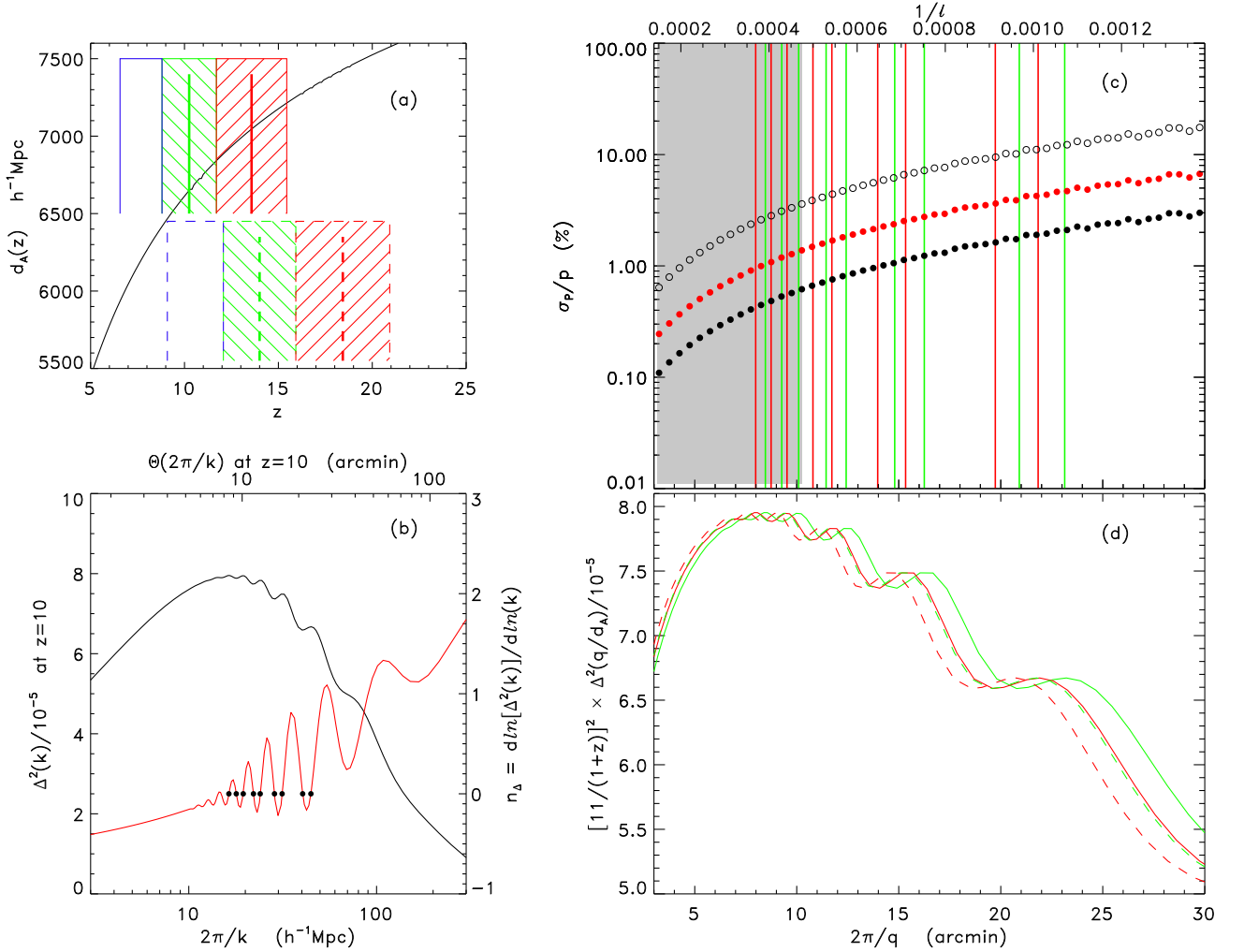}
\caption{\small (a) Solid line shows $d_A$ span vs $z$ for flat Universe with $\Omega_m=0.3,\Omega_\Lambda=0.7$. The span of $z_{\rm Ly}$ over Y, J, and H {\it Euclid} filters is shown in blue, green, and red; vertical lines correspond  to the central wavelength of each filer. Upper regions correspond to the Lyman-break at Ly-$\alpha$ and lower (dashed) at Ly-continuum. (b) Solid line shows $\Delta^2(k)$ evaluated with CMBFAST at $z=10$. Red line and right axis show the spatial spectral index, $n_\Delta$, of $\Delta^2(k)$; solid dots mark extrema of $\Delta^2(k)$. For Harrison-Zeldovich regime $n_\Delta=3$ which is reached at larger scales. (c) Relative accuracy for probing the tomographically measured power at each angular frequency. Open, red and black filled circles correspond to selecting a $21^\circ\times21^\circ$ deg field, 1yr and full {\it Euclid} Wide Survey areas. Shaded region shows the angular scales covered by one {\it Euclid} detector. Vertical lines mark the BAO extrema in $\Delta^2$ for J--Y (green) and H--J tomographic maps. (d) $\Delta^2(q/d_A)$ at the redshifts marked with vertical lines in (a). 
}
\label{fig:fig1}
\end{figure}

We seek to isolate the power, $P_{\Delta z}$, from luminous sources between
 $z_{\rm Ly}(\lambda_1)$ and $z_{\rm Ly}(\lambda_2)$.  
We rewrite (\ref{eq:limber_2}) to isolate CIB fluctuation at $z_{\rm Ly}(\lambda_1)<z<z_{\rm Ly}(\lambda_2)$:
\begin{equation}
\frac{q^2P_{\Delta z}(q)}{2\pi}= \left[\frac{q^2}{2\pi}(P_2-\frac{P_{12}^2}{P_1})\right]_{\rm data} + \frac{q^2}{2\pi}P_{\rm sys}
\label{eq:p_df}
\end{equation}
where the first rhs term is fully given by the data and the last term is driven by incoherence of the sources at the two adjacent bands which occupy the {\it same} span of redshifts $z<z_{\rm Ly}(\lambda_1)$:
\begin{equation}
\frac{q^2}{2\pi}P_{\rm sys}\;\; =\;\; \left[\frac{{\cal C}_{12}(q, z<z_{\rm Ly}(\lambda_1))-1}{{\cal C}_{12}(q,z<z_{\rm Ly}(\lambda_1))}\right]\; \times\; \left[\frac{q^2}{2\pi}\frac{P_{12}^2}{P_1}\right]_{\rm data}\leq0.
\label{eq:systematic}
\end{equation}
The subscript ``data" refers to directly measurable quantities. $P_{\rm sys}\leq 0$ because $0 \leq {\cal C}\leq 1$. Thus, the measurable quantity $\Delta P_{\rm obs}\equiv (P_2-\frac{P_{12}^2}{P_1})$ sets an {\it upper} limit on the CIB fluctuations arising at $z_{\rm Ly}(\lambda_1)<z<z_{\rm Ly}(\lambda_2)$. This methodology has already been successfully applied to deep {\it Spitzer} data,
leading to interesting upper limits on emissions
at $30\lsim z\lsim 40$ (Kashlinsky et al. 2015 - K15).

{\it Euclid}'s NISP instrument will have three near-IR filters which are referred below as $Y, J, H$ in order of increasing central wavelength: 1.056, 1.368, 1.772 \micron. Each band will be available for evaluating CIB fluctuations. Figure \ref{fig:fig1}a shows that the 
{\it currently} envisaged Y, J, and H filters (K. Jahnke, 08/2015, private communication) isolate emissions over narrow ranges, $\sim 5-7\%$, in $d_A$. 
We assume that at these epochs the power spectrum of the emitting sources is proportionally related to the  underlying $\Lambda$CDM 
one: $\Delta^2(q/d_A)=b^2(z)\Delta^2_{\Lambda CDM}(q/d_A)$ with $b$ being the bias factor, since the relevant angular scales subtend tens of comoving Mpc where density field was highly linear. Because the 
procedure isolates a narrow shell in $d_A(z)$ around $d_0$, the comoving angular distance to the central filter wavelength, we further expand $\Delta^2_{\Lambda CDM}(q/d_A) \simeq \Delta^2_{\Lambda CDM}(qd_0^{-1})\{1-n_\Delta(qd_0^{-1})[\frac{d_A(z)-d_0}{d_0}]\}
$, where $n_\Delta(k)\equiv d\ln\Delta^2_{\Lambda CDM}(k)/d\ln k$ is the spatial spectral index of the $\Lambda$CDM template shown in Figure 
\ref{fig:fig1}b. Further noting that $d_A-d_0\simeq cH_0^{-1}(z-z_0)/E(z)$ we write the power
from sources over the narrow range of epochs defined in Figure \ref{fig:fig1}a as:
\begin{equation}
\frac{q^2P_{\Delta z}}{2\pi}\simeq \Delta^2_{\Lambda CDM}(qd_0^{-1}) \times 
\int_{z_{\rm Ly}(\lambda_1)}^{z_{\rm Ly}(\lambda_2)} \left(\frac{dF_{\lambda_2^\prime}}{dz}\right)^2b(z)^2 \left[E(z)
-n_\Delta(qd_0^{-1})\frac{R_H}{d_0}  (z-z_0)\right]dz.
\label{eq:p_df_1}
\end{equation}
This relates $P_{\Delta z}$ to the underlying parameters over the narrow range of $z$. Figure \ref{fig:fig1}b
shows the underlying shape of $\Delta^2(k)$ with the BAO oscillations being prominent at potentially measurable levels. 
The integration in (\ref{eq:p_df_1}) represents a convolution of the 
BAO spectrum over the redshift range defined by the wavelengths $\lambda_1$ and 
$\lambda_2$. For $\lambda_1$ and $\lambda_2$ set by the red edges of the 
Y, J, and H filters, this convolution alters the amplitude of the power 
spectrum by $<1\%$ if $dF/dz$ is constant across the band. 


\section{Analysis configuration and uncertainties}

After 6.25 years, {\it Euclid}'s Wide Survey (EWS) will cover 15,000 deg$^2$ at Y, J, H to $m_{\rm AB}\simeq 25$ (3$\sigma$); Euclid's Deep Survey (EDS) will cover (non-contiguously) 40 deg$^2$ two magnitudes deeper. The derived Y, J, and H CIB maps will be used in the tomographic reconstruction, [J--Y] and  [H--J], isolating 
populations over $\delta d_A\ll d_A$ (Figure \ref{fig:fig1}a).
Equation (\ref{eq:p_df_1}) shows that scatter around the expected $\Lambda$CDM template at an effective $d_A$ caused by the finite range in $z$ probed in each 
of the tomographic constructions, will be generally small and even further reduced around the extrema of the template (Figure \ref{fig:fig1}b,c). Figure \ref{fig:fig1}d shows the expected angular spectra of the CIB fluctuations with BAO structures.


There are two criteria here: 1) does the measured signal fit the expected  $\Lambda$CDM template?; and if it does 2) how well can the physically important parameters of that template be measured from the data?

The first criterion requires sampling the power with sufficient angular resolution, say $\Delta \theta \sim 0.5'$,
at scales of $5'\leq \theta = 2\pi/q\lsim 25'$ where the BAO structure lies. In terms of angular frequency, this means
$\Delta q = 2\pi \Delta \theta / \theta^2 \approx \frac{1}{200}$~amin$^{-1}$. The frequency resolution, $\Delta q$
is set by maximum size, $\Theta_0$, of the region being analyzed: $\Delta q = 2\pi /\Theta_0$. Therefore, to achieve 
sufficient sampling to resolve the BAO structures, the analysis requires regions that are 
$\Theta_0 = 2\pi/(5\times 10^{-3}\ {\rm amin}^{-1}) \approx 21\arcdeg$ in size. 
Masking out resolved sources introduces small-scale coupling increasing $\Theta_0$ further, although this effect would be small when probing power at $\Delta\theta\simeq 0.5'$. If so, the large value of $\Theta_0$ requires use of  
EWS rather than EDS, potentially necessitating spherical harmonic analysis. However, this conservative estimate can
be relaxed if the data show BAO structure dominated by emissions from one effective $z$. In that case, EDS' smaller fields, with lower CIB from remaining galaxies,  may also prove useful. To sample each of $N_a(\sim4)$ acoustic peaks with $N_{\rm sampling}>2$ points/peak between $2\pi/25' \leq q \leq 2\pi/5'$ will lead to $\Delta q =2\pi(\frac{1}{5'}-\frac{1}{25'})(N_aN_{\rm sampling})^{-1}\simeq 2\pi/250'$,  or $\Theta_0\sim 4^\circ$ for $N_aN_{\rm sampling}=40$.


%

Once the auto- and cross-power spectra are measured between channels 1 and 2 (Y and J, J and H) the quantity to be determined is $\Delta P_{\rm obs}$, which is to be compared to $P_{\rm \Delta z}$.  The relative accuracy on this quantity can be evaluated as sampling (cosmic) variance or $\sigma_P/P= \sqrt{6/
N_q}$ with $N_q$ being the number of independent Fourier elements that go into determining the power at each $q$ (K15). The amplitude of the $\Lambda$CDM fit, $F_0$, can be evaluated iteratively from 
smaller angular scales in conjunction with $d_A$ determined from larger scales; we assume that scales $2\pi/q>1'$ are left to evaluate the effective $d_A$ and its uncertainty from linear 
least-squares regression. We model the observationally determinable $q^2\Delta P_{\rm obs}/(2\pi)$ as $M_i=T(q_i/d_A)+M_{\rm sys}(q_i)$ where $M_{\rm sys}$ is given by Equation (\ref{eq:systematic}) and $T(k)=F_0^2\Delta^2_{\Lambda CDM}(k)$.  This theoretical fit model is inherently highly-nonlinear. Once data ($\vec{D}$) are available, Markov chain 
processing may be used to evaluate the range of cosmological parameters and their uncertainties. To estimate 
the uncertainty on $d_A$ we 1) note the narrow width of the surface where the sources contributing to $P_{\Delta z}$ are located, 2) make an initial guess $d_A=d_0$, and 3) then linearize the model $M_i \simeq T(q_i/d_0)[1-n_\Delta(q_i/d_0) \epsilon_d]+M_{\rm sys}(q_i)$ with $\epsilon_d=(d-d_0)/d_0$. The result is then obtained by minimizing $\chi^2=\langle \frac{(D_i-M_i)^2}{\sigma_i^2}\rangle$ with $\sigma_i$ shown in Figure \ref{fig:fig1}c. 
The statistical uncertainty on the distance determination from $\partial \chi^2/\partial \epsilon_d=0$ is then:
\begin{equation}
\sigma_\epsilon^2 = \sum_i (\frac{\partial \epsilon_d}{\partial D_i})^2\sigma_i^2 = \frac{6}{\sum_i [n_\Delta(q_i/d_0)]^2N_{q,i}}
\label{eq:sigma_d}
\end{equation}
Assuming $1'<2\pi/q <25'$ leads to $d_A$ being probed 
with a relative accuracy of 1.25\% from a single field of 400 deg$^2$ (as in Figure \ref{fig:fig1}c). Scales $1'-10'$, which lie entirely within
one detector, account for about 90\% of this. In larger EWS areas the relative uncertainty will
improve $\propto 1/\sqrt{{\rm area}}$. Systematic uncertainty is given by the last term in $M_i$; its amplitude is illustrated in the following section.






Diffuse light fluctuations  from remaining galaxies and foregrounds may contribute, although  at {\it Spitzer} wavelengths they are much lower than the CIB fluctuation (KAMM1, HRK12, K12); additionally, they should be highly coherent. We discuss the contributions from the known remaining galaxies in the EWS in the next section. The estimated power from the zodiacal light and diffuse Galactic light (DGL) 
are at similar levels to that of the 
remaining known galaxies (KAMM1, AKMM, and K15). 
These foregrounds will affect the measured $\Delta P_{\rm obs}$ at a 
similar level as the remaining known galaxies, provided they have comparably 
high coherence. The Kelsall et al. (1998) zodiacal light model 
suggests a color gradient of $\lesssim0.1\%$ per degree
is present between the {\it Euclid} bands. Measured color variations 
of the DGL can reach factors of 2 between different locations 
(Ienaka et al. 2013), indicating 
relatively low coherence and a stronger contribution to $\Delta P_{\rm obs}$
on large angular scales.

%
\section{Modelling astrophysical applications}

This method applied to  large contiguous areas of EWS can 1) accurately isolate the history of emissions at $10 \lsim z \lsim 20$ and 2) measure the BAO at those epochs. 
The discussion above shows that if the
CIB produced by pre-reionization sources is high enough to be directly measurable with {\it Euclid}, both of these goals can be achieved. In this section we illustrate the feasibility of these goals with a specific model for high-$z$ evolution in the presence of CIB from known galaxy populations that will be remaining in the {\it Euclid} data. The high-$z$ modeling, while consistent with all current data, is used for illustrative, not predictive purposes.

The model adopted here for illustrative purposes is IMF500 described in detail in Helgason et al. (2015). We assume that dark matter halos collapse to form the stars with a fixed efficiency $f_*$ until $z_{\rm end}$ with the rate of collapsed dark halos fixed by the power spectrum evolution in the $\Lambda$CDM model; the levels of the CIB left behind by these sources are varied via $f_*$. For 
$f_*=0.04$ ($z_{\rm end}=10$) ($f_*=0.03$ for $z_{\rm end}=8$) the model reproduced the source-subtracted CIB fluctuation measured in the {\it Spitzer} data by K12; lowering $f_*$ would reduce the CIB fluctuation power $\propto f_*^2$. The CIB fluctuation of remaining known galaxies in EWS is calculated using the HRK12 reconstruction technique for three limits: the default reconstruction, which is supported by the later measured {\it Spitzer} deep counts data (Ashby et al. 2013, 2015), and two extreme limits of possible extrapolation of the observed luminosity functions, termed the high-faint-end (HFE) and low-faint-end (LFE) limits. 
The HRK12 empirical reconstruction employs an assembled extensive database of galaxy luminosity functions 
spanning a wide range of wavelengths, redshifts and luminosities; it was demonstrated 
to agree well with both galaxy counts from visible to near-IR and numerical 
modeling of galaxy evolution.
\begin{figure}[h!]
\hspace{-15mm}
\includegraphics[width=7in]{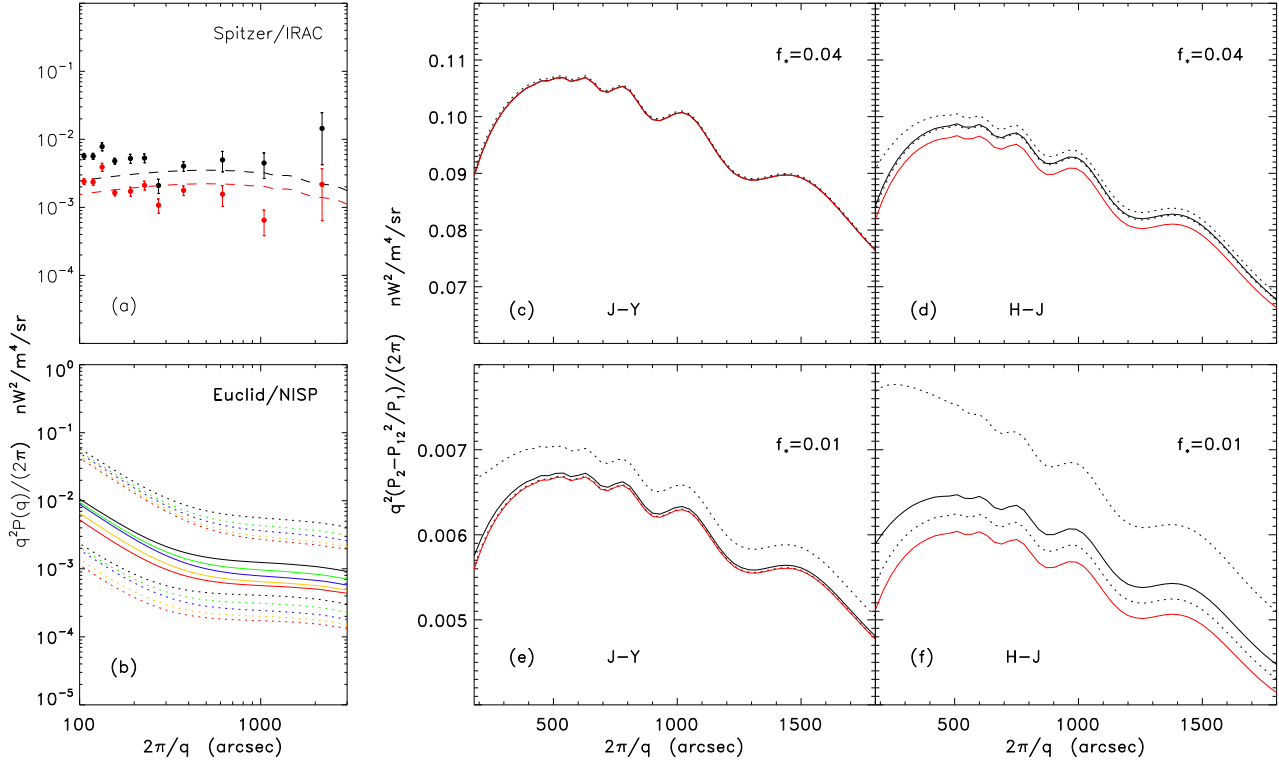}
\caption{\small (a) {\it Spitzer} data at 3.6 (black) and 4.5 (red) \mic\ (K12). IMF500 model described in the text is shown with dashed lines at $f_*=0.04$
for $z_{\rm end}=10$ ($f_*=0.03$ for $z_{\rm end}=8$). (b) HRK12 reconstructed CIB fluctuation from galaxies remaining unresolved in NISP data. Black, blue, red, green, yellow correspond to Y, J, H and J$\times$Y, H$\times$J configurations. Solid lines of each color correspond to the default reconstruction and dashes show the range of HFE to LFE extrapolations of the luminosity function data. (c)-(f)  reconstructed $q^2\Delta P_{\rm obs}/(2\pi)$ from the high-$z$ model plus the remaining galaxies at levels shown in (b): solid lines for the default reconstructions and dotted lines for the HFE (upper) and LFE limits. Red lines show the underlying CIB fluctuation produced at $z_1< z < z_2$.}
\label{fig:fig2}
\end{figure}

\subsection{Isolating emissions at $z\gsim 10$}

Figure \ref{fig:fig2}a shows the {\it Spitzer}-based source-subtracted CIB fluctuations (K12) at 3.6 and 4.5 \mic\ compared with the high-$z$ sources in the above modeling. Presently there is no direct evidence that the measured fluctuations originate at high-$z$, although they appear uncorrelated at any significant level with diffuse visible-band light from sources at $m_{\rm AB}>28$ (Kashlinsky et al. 2007c). Providing such evidence directly from direct Lyman-break CIB measurements requires eliminating sources to much fainter limits than is possible with current experiments and will be achievable with {\it JWST} and {\it Euclid} (K15). We model the possibility of only a fractional power contributed to the measured CIB by lowering  $f_*$; e.g. if $f_*=0.01$ only $\sim6\%$ of the measured CIB power (Figure \ref{fig:fig2}a) arises at high $z$.  

To estimate how well we would recover the emissions from the $z$-interval defined by each adjacent filter pair, we construct each of the auto- and cross-power spectra to combine the contributions from the high-$z$ model populations and known galaxies remaining in the EWS. The CIB fluctuation contributions from the known galaxies remaining in each {\it Euclid}/NISP filter for EWS are shown in Figure \ref{fig:fig2}b for the default, HFE and LFE reconstructions.
For given $f_*$, we construct the quantity $\Delta P_{\rm obs}$ for each of the J--Y and H--J configurations and compare it with the directly computed $P_{\Delta z}$ due to emissions over the width of the J and H filters (Figure \ref{fig:fig1}a).  

Figure \ref{fig:fig2}c-f compares the CIB fluctuations recovered by the proposed tomography method (black lines) with the true signal produced over the given $\Delta z$ (red solid). The figure illustrates that if the entire CIB signal discovered in {\it Spitzer}-based measurements (KAMM1, KAMM2, K12) originates at high $z$, this method reconstructs emission history with high accuracy (better than 6\% for this illustrative model). Even if only a fraction ($\sim 6\%$ for this model) of the signal comes from high $z$, the accuracy remains interestingly high (better than $\sim 20\%$  for this model). 

\subsection{Probing BAO at $z\gsim 10$}

\begin{figure}
\centering
\includegraphics[width=4in]{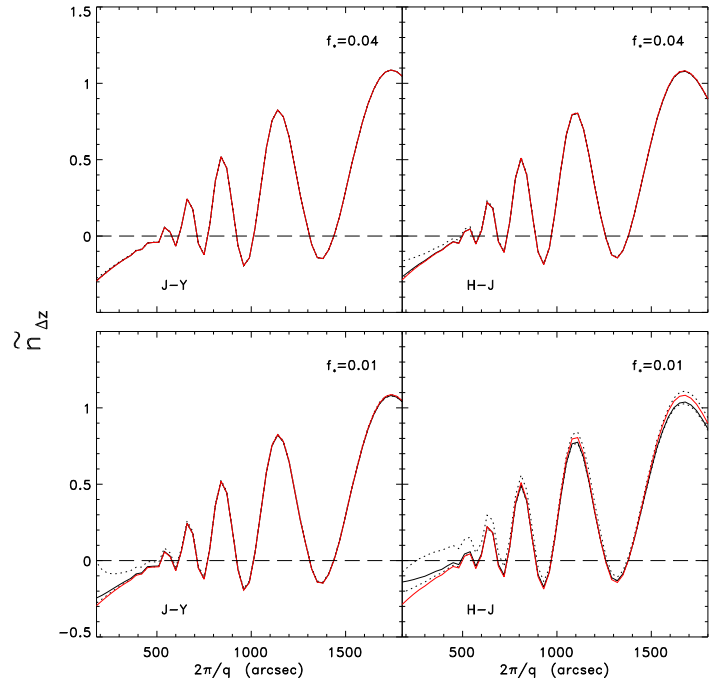}
\caption{
The effective CIB spatial spectral index at $[z_1,z_2]$ from $\Delta P_{\rm obs}$ (black) vs $P_{\Delta z}$ (red). Same line notation as in Figures \ref{fig:fig2}c-f. 
}
\label{fig:fig3}
\end{figure}

While there appears a small upward bias in the emissions' amplitude (typically $\lsim 10\%$) estimated by this method, the recovered angular shape of the CIB fluctuations (black lines) is in very good agreement with the true angular profile (red). This is shown in Figure \ref{fig:fig3} which plots the resultant 2-D CIB power index $\tilde{n}_{\Delta z} \equiv \partial{\ln [q^2\Delta P_{\rm obs}]}/\partial{\ln q}$. Even for the highly pessimistic case of HFE reconstruction and $f_*=0.01$, the ratio of the black to red lines remains constant at large scales of $\gsim 10'$. This argues for good prospects of BAO measurement at these epochs from the application of the Lyman tomography method to the upcoming {\it Euclid} CIB maps which will provide an important consistency
check of the standard cosmological model.

\begin{figure}
\centering
\includegraphics[width=7in]{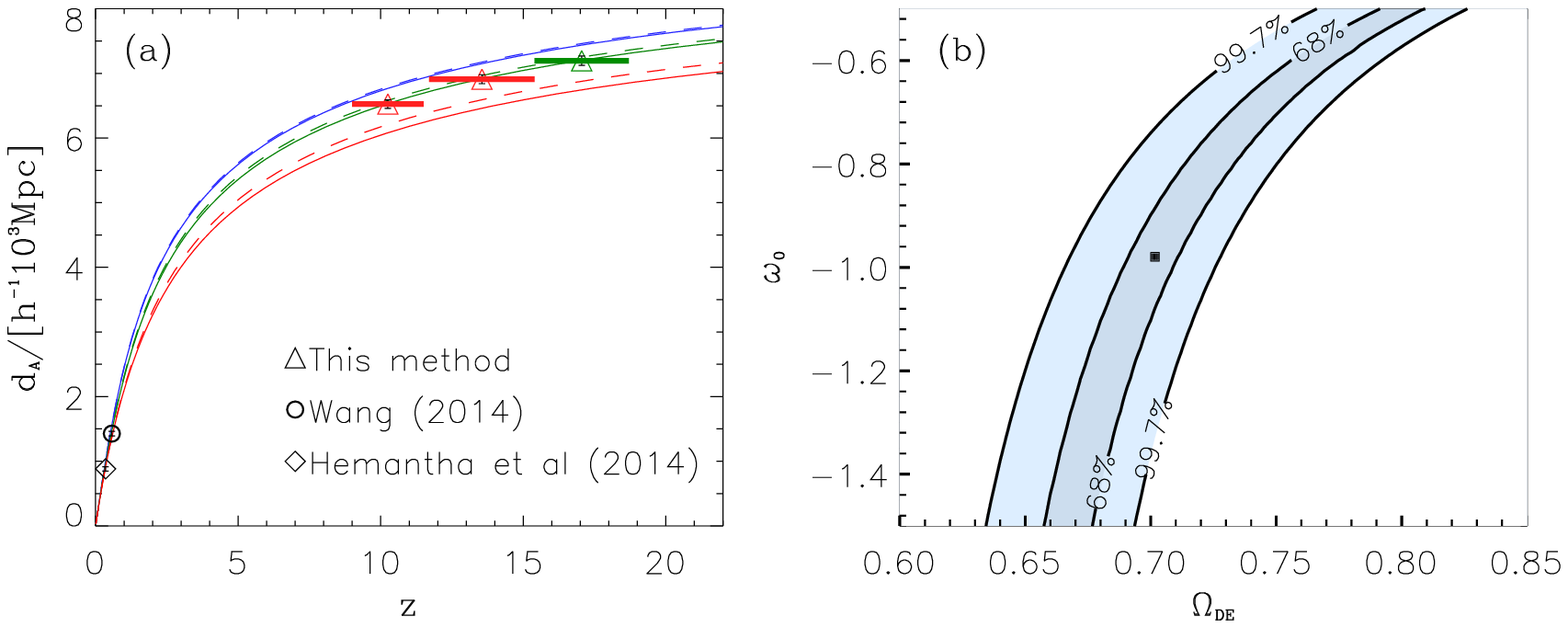}
\vspace*{-4cm}
\caption{
(a) $d_A$ with DE evolving according
to the CPL EoS parametrization.  Blue-top, green-middle
and red-bottom lines correspond to $\omega_0=[-1.4,-1,-0.6]$. Solid/dashed lines denote $\omega_a=[-0.03,-0.3]$.
Diamond and circle symbols correspond to current data and triangles to this
methodology applied to {\it Euclid}'s (red) CIB and potentially an additional {\it WFIRST} (green) measurement; horizontal bars span
the uncertainty in $z$ for each {\it Euclid} filter configuration.
(b) Confidence contours obtained from BAO measured at
the three different redshifts given in (a).
The black square corresponds to the best fit value; in this case
the fiducial concordance model.
}
\label{fig:fig4}
\end{figure}

 The sound horizon at the end of the radiation drag is a (BAO) scale imprinted
in $P_{3D}$. BAOs allow us to measure the angle subtended  by the scale, which is directly related to $d_A$. 
The power spectra of Figure~2 can be used to estimate the angular size 
subtended by the measured sound horizon scale, $r_s=144.81\pm 0.24$~Mpc 
(Planck Collaboration 2014) as described
in Percival et al. (2010). With the proposed tomography we can potentially determine $d_A(z)$ to
$\lsim 1\%$ accuracy, but with a systematic uncertainty on $z$ due to the 
Lyman-break position. Eisenstein et al. (1998) proposed to 
constrain cosmological parameters using BAO measurements (e.g. Weinberg et al. 2013).
Figure~\ref{fig:fig4}a plots $d_A$ for different values of the DE equation-of-state parametrization 
$\omega_{\rm DE}(z)=\omega_0+\omega_a z/(1+z)$ (Chevalier \& Polarski 2001, Linder 2003).
Triangles with red horizontal bars correspond to the proposed Lyman tomography CIB analysis of EWS. The
vertical error bars (barely noticeable) correspond to 1\% relative errors.
Horizontal lines represent the redshift span of the 
CIB sources given in Figure~\ref{fig:fig1}a at the Ly$\alpha$-break.
The data will constrain cosmological parameters at $z\gsim 10$.
{\it WFIRST} \citep{wfirst}
will carry out complementary observations to those of {\it Euclid}.  In Figure \ref{fig:fig4}a the 
green triangle shows 
 the advantage of adding 
a {\it WFIRST} filter covering $2-2.4~\mu$m
that could provide the BAO scale at an additional $z$ with this methodology.
For comparison, we plot data from
Hemantha et al. (2014) and Wang (2014) at low $z$.
In Figure~\ref{fig:fig4}b we derived the confidence contours on the
parameters $(\Omega_{\rm DE},\omega_0)$, assuming $\omega_a=0$.
The width of the contours is dominated by the finite span of $z$.
While the method may not constrain those parameters as well as 
other techniques, it extends the BAO regime to hitherto unprobed 
$z$ and provides an important self-consistency check. 
Alternatively, if the cosmological model is assumed, one can compare 
the measured $d_A$ with the expected value at different $z$ 
to determine the effective redshift of the sources that contribute 
to each of the three measurements of Figure~\ref{fig:fig4}a. A 0.2\% (1\%) relative error 
on the BAO angular diameter distances allows determination of $d_A$ with an 
accuracy $\Delta z=0.09,0.15,0.2\; (0.45,0.75,1)$, respectively.
These uncertainties can be further reduced if the cross-correlation
of CIB fluctuations and CMB temperature anisotropies, which are also
potentially sensitive to BAOs (Atrio-Barandela \& Kashlinsky 2014), is measured.

%
%

\acknowledgements
We thank Jason Rhodes for comments, Alexander Vassilkov for discussion of statistical treatment and  NASA/12-EUCLID11-0003 ``LIBRAE: Looking at Infrared Background Radiation Anisotropies with Euclid" for support. FAB acknowledges the Ministerio de Educacion y Ciencia project FIS2012-30926 and KH from EU's 7th Framework Programme (FP7-PEOPLE-2013-IFF).


\end{document}